\documentstyle[prb,aps]{revtex}

\begin{document}
\twocolumn[\hsize\textwidth\columnwidth\hsize\csname 
@twocolumnfalse\endcsname

\title{Current Density and Reversible Magnetization of 
HgBa$_2$Ca$_2$Cu$_3$O$_x$ Superconductors 
Containing Randomly Oriented Columnar Defects}
\author{J. R.Thompson,$^{a,b}$ J. G. Ossandon,$^c$ L. Krusin-Elbaum,
$^d$ H. J. Kim,$^b$ D. K. Christen,$^a$ K. J. Song,$^b$ 
K. D. Sorge,$^b$  and J. L. Ullmann$^e$}
\address{$^a$ Oak Ridge National Laboratory, Oak Ridge, 
Tennessee, 37831-6061 USA \\
$^b$ Department of Physics, University of Tennessee, 
Knoxville, Tennessee 37996-1200 USA \\
$^c$ Department of Engineering Sciences, University 
of Talca, Curico, Chile \\
$^d$ IBM Watson Research Center, Yorktown, New York, 10598
USA \\
$^e$ Los Alamos National Laboratory, Los Alamos, New 
Mexico, 87545 USA}

\date{29 April 2002}
\maketitle

\begin{abstract}
Bulk polycrystalline 
HgBa$_2$Ca$_2$Cu$_3$O$_x$ materials were 
irradiated with 0.8~GeV protons to form randomly 
oriented columnar defects, by induced fission of 
Hg-nuclei.  Proton fluences from 0 to $35 \times 
10^{16}$ cm$^{-2}$ were used to install defects 
with area densities up to a ``matching field" of 
3.4 Tesla.  Studies were conducted on the 
dependence of the equilibrium magnetization and 
the intragrain persistent current density on 
temperature and applied magnetic field, at 
various defect densities.  The magnetization was 
modeled using London theory with the addition of 
vortex-defect interactions. \\

Corresponding author:  
James R. Thompson
Oak Ridge National Laboratory
PO Box 2008
Oak Ridge, TN 37831-6061 USA
Phone: 865-574-0412; fax: 865-574-6263; email:  
JRT@UTK.EDU

\end{abstract}

\vskip1pc] \narrowtext

\section{Introduction}
For strong pinning of vortices in high-$T_c$ 
superconductors, some form of correlated disorder 
generally is most effective. Most widely studied 
have been columnar defects, which typically are 
formed by irradiation with energetic heavy ions; 
such particles are highly ionizing and create 
tracks of amorphous material along their path.\cite{Civale92}   
Many morphologies have been investigated, ranging 
from the familiar parallel tracks to inclined 
defects, crossed arrays with ``designer splay," 
etc.  One problem with heavy ions is their 
limited range, typically a few tens of 
micrometers.  To overcome the limited range of 
heavy ions, Krusin-Elbaum et al.\cite{LKE94} 
demonstrated an 
indirect formation of columnar defects using 
deeply penetrating 0.8~GeV protons.  In this 
process, an incident proton is absorbed by a 
heavy nucleus (Hg) within the material and 
excites it into a highly energetic state.  As a 
consequence, the nucleus breaks into two 
particles with similar mass, each having $\sim$ 100~MeV 
energy.  The fission fragments form randomly 
oriented columnar defects (CD's) deep within the 
superconductor.  In a different approach to 
create pins deep inside a material, Weinstein and 
coworkers have created CD's by introducing 
$^{235}$U as a dopant into the superconductor, 
which is induced to fission by irradiation with 
thermal neutrons.\cite{Schultz98}
 
Vortex pinning by proton-generated fission 
defects was first demonstrated in Bi-cuprate 
materials.\cite{LKE94,Safar95} 
In fact, the proton-based fission 
process is generally applicable in cuprate 
superconductors that contain a sufficient density 
of heavy nuclei.\cite{Thompson97} 
High-$T_c$ materials containing 
randomly oriented CD's exhibit a variety of 
interesting physical phenomena, e.g., temperature 
independent quantum tunneling of vortices in Bi-2212 
materials.\cite{Thompson99}
In studies of a series of Hg-cuprates 
containing 1, 2, and 3 adjacent CuO 
layers, it was shown\cite{LKE98}
that sufficiently high 
superconductive anisotropy can lead to a 
rescaling of the splayed landscape of random 
CD's.  As a consequence, the pinning array and 
the applied field are `refocused' toward the 
crystalline $c$-axis, even in polycrystalline 
materials. This is a useful feature that we will 
use when interpreting present studies of the 
equilibrium magnetization.  

In this paper, we describe how the addition 
of random CD's changes the superconductive 
properties of HgBa$_2$Ca$_2$Cu$_3$O$_x$ 
superconductors.  Specific topics include the 
intragrain persistent current density $J$, as 
obtained from the magnetization; establishment of 
the optimum defect density; and analysis of the 
changes in the mixed state reversible (or 
equilibrium) magnetization resulting from 
interactions between vortices and randomly 
oriented columnar defects.

\section{Experimental Aspects}
Samples~for~study~were bulk polycrystalline 
HgBa$_2$Ca$_2$Cu$_3$O$_x$ materials (Hg-1223) 
containing sets of 3 adjacent oxygen-copper 
layers.  Small pieces, typically 30~mg mass and 
$\sim$ 1 millimeter thickness, were all cut from 
the same pellet.  The samples were irradiated at 
room temperature in air with 0.8~GeV protons at 
the Los Alamos National Laboratory.  Proton 
fluences $\phi_p$ were 0, 1.0, 3.2, 10, 19, and 
$35 \times 10^{16}$~protons/cm$^2$, as determined 
from the activation of Al dosimetry foils. The 
resulting density of fission events is $N/V = 
\phi_{p} \sigma_{f} n_{Hg}$.  Here $n_{Hg}$ is 
the number density of Hg nuclei and $\sigma_{f} = 
\sim 80$ millibarns is the cross-section for 
inducing a prompt fission of Hg nuclei.  Each 
fission produces one CD, so $N/V$ is the volume 
density of defects, also.  One can convert this 
into an approximate area density of CD's by 
multiplying by the track length $\sim$ 8~$\mu$m, 
and the area density can be reexpressed in units 
of a matching field $B_{\Phi}$ by multiplying by 
the flux quantum $\Phi_0$. Resulting values for 
the defect density are $B_{\Phi}$ = 0, 0.1, 0.3, 
1.0, 1.9, and 3.4 Tesla, respectively. The 
fission process is random in direction, giving 
randomly oriented CD's.  Transmission electron 
microscopy (TEM) studies of both Bi-2212 and 
Hg-cuprates\cite{LKE97} have demonstrated the presence of 
randomly oriented columnar defects in these high-$T_c$ materials. 

The superconductive properties of the virgin 
and processed samples were investigated 
magnetically.  A SQUID-based magnetometer (model 
Quantum Design MPMS-7), equipped with a high 
homogeneity 7~T superconductive magnet, was used 
for studies in the temperature range 5--295~K, in 
applied fields to 6.5~T.  The superconductive 
transition temperatures $T_c$ were measured in an 
applied field of 4~Oe (0.4~mT) in zero-field-cooling 
(ZFC) and field-cooling (FC) modes.   The 
resulting values for the onset temperature $T_c$ 
are 133, 134, 132.3, 132, 131, and 129.3~K, 
respectively.  Values for the Meissner (FC) 
fraction $-4 \pi M/H$ lie in the range of 40--
50\%, except at the highest fluence where the 
fractional flux expulsion was 29\%. 

The isothermal magnetization $M$ was 
measured as a function of applied magnetic field.  
Below $T_c$ and below the irreversibility line, 
the magnetization was hysteretic due to the 
presence of intragrain persistent currents.  From 
the magnetic irreversibility $\Delta M = 
[M(H\downarrow) - M(H\uparrow)]$, the persistent 
current density $J$ was obtained using the Bean 
critical state relation $J \propto \Delta M/r$, 
where $r \approx 4$~$\mu$m is the mean grain 
radius. Measurements of the background 
magnetization in the normal state above $T_c$ 
were used to correct the data in the 
superconducting state, in order to obtain the equilibrium 
magnetization $M_{eq}$.  Above the 
irreversibility line where $\Delta M = 0$, this 
process yields $M_{eq}$ directly; near the 
irreversibility line where $\Delta M$ is small, 
we obtain $M_{eq}$ from the (background-corrected) 
average magnetization $ 
[M(H\downarrow) + M(H\uparrow)]/2$, as 
illustrated later.

\section{Experimental Results}
We begin with a discussion of the effects of 
randomly oriented columnar defects on the 
irreversible properties of the Hg-1223 
superconducting materials.  The addition of 
correlated disorder increases the pinning of 
vortices, often quite significantly.  An example 
is contained in Fig. 1, which shows the 
magnetization $M(H)$ at 60~K vs magnetizing field 
$H$, for various defect densities.  With 
increasing fluence, the ``hysteresis loops" 
increase in width and become symmetric about the 
$M = 0$ axis.  (The asymmetry for the virgin 
sample, coupled with the fact that the decreasing 
field branch lies near the $M = 0$ axis, suggests 
that surface barriers \cite{Sun94,Lewis95,Kim95} 
contribute to the 
observed hysteresis in this case.  It is 
interesting to note that a relatively low dosage 
of CD's, $B_{\Phi} = 0.1$~T, symmetrizes the 
$M(H)$ loop significantly; in particular, the 
magnitude of $M$ in the decreasing field branch 
is much larger.  These features imply that the 
addition of CD's suppresses the surface barrier, 
as recently discussed by Koshelev and Vinokur.\cite{Koshelev01})

Using the Bean model, one may obtain the 
persistent current density $J(H,T)$.  For these 
weakly linked polycrystalline materials, the 
magnetization reflects the {\em intragranular} 
current density.  Some results of this analysis 
are shown in Fig. 2 and Fig. 3 as plots of $J$ 
versus temperature $T$, in applied fields of 0.1 
and 1 Tesla, respectively.  The enhancement in 
$J$ is modest in low fields; this can be 
attributed to the presence in the virgin sample 
of both naturally occurring defects that provide 
some pinning and the likely influence of surface 
barrier effects, as already noted. In higher 
fields, the contribution of the random CD's is 
more apparent and $J$ is enhanced by about an 
order-of-magnitude.  For high-$T_c$ materials 
with these angularly dispersed defects, one 
generally achieves the maximum $J$ at some defect 
density $B_{\Phi}$ near 0.2--2~T.  This 
optimization is shown by the insets in Figs. 2 
and 3.  For an applied field of 0.1~T (Fig. 2 
inset), $J$ at 100~K is largest for $B_{\Phi} = 
0.1$--0.3~T.  For the second example with a field 
of 1~T (Fig. 3 inset), $J$ at 60~K is largest for 
$B_{\Phi}$ near 2~T.  Qualitatively, the optimum 
defect density in each case is comparable with 
the vortex density, i.e., about $2\times$ larger. 
At higher defect densities, $J$ decreases.  This 
may be due in part to the presence of additional 
CD's that help to initiate the hopping of 
vortices to nearby empty pinning sites.  A 
second, potential influence is a suppression of 
$T_c$; however, the change in $T_c$ is small and 
these measurements are far from $T_c$, so this 
contribution is expected to be small. A third and 
sizable effect is an irradiation-induced 
suppression of the vortex line energy and pinning 
energy, due to changes in the London penetration 
depth; this will be discussed in a following 
section.  The optimum defect density depends 
weakly on the temperature as well.

Next we consider the equilibrium 
magnetization in the mixed state.  Figure 4 
illustrates the method used to obtain $M_{eq}$ by 
plotting the background-corrected experimental 
magnetization for the virgin material (open 
squares) versus applied magnetic field $\mu_0 H$ 
on a logarithmic scale.  The closed symbols show 
the average $M$, which provides a very good 
approximation to $M_{eq}$ when the hysteresis is 
small.  The solid line is a fit to conventional 
London theory, which provides that $M_{eq}$ is 
directly proportional to $(1/\lambda_{ab}^{2}) 
\times \ln(\beta B_{c2}/H)$, where $\lambda_{ab}$ 
is the in-plane London penetration depth, $\beta$ 
is a constant of order unity, and $B_{c2}$ is the 
upper critical field.\cite{Kogan88} The straight line fit in 
Fig. 4 shows that the London $\ln(H)$ logarithmic 
variation describes the field dependence well. 
Results for $M_{eq}$ in the virgin Hg-1223 
material at other temperatures are presented in 
Fig. 5.  One sees from the linear regression 
lines that the London field dependence is 
followed over a wide range of fields and 
temperatures.  Deviations from linearity occur at 
low fields when $H$ approaches $H_{c1}$ where 
simple London theory is not valid, and at low 
temperatures where the materials are sufficiently 
hysteretic that the averaging procedure 
illustrated in Fig. 4 is not valid. The slopes of 
the curves in Fig. 5 provide values for the 
London penetration depth $\lambda_{ab}(T)$ in the 
virgin material.  These results will be compared 
and contrasted with values deduced for the 
irradiated Hg-1223. 

We now address the question of how the 
addition of randomly oriented columnar defects 
modifies the equilibrium magnetization in the 
mixed state.  The qualitative effects of the CD's 
are illustrated in Fig. 4, which includes data 
for $M_{eq}$ at 77~K for irradiated samples with 
defect densities $B_{\Phi} \approx 1.0$ and 
3.4~T.  Comparing these data with the virgin 
curve shows that the CD's reduce the magnitude of 
the equilibrium magnetization considerably, 
especially at lower fields.  In addition, they 
generate a pronounced deviation from the 
``standard" London field dependence, giving 
$M_{eq}$ an ``S"-like dependence on field.   
This ``anomalous" behavior has been observed 
previously in cuprates containing {\em parallel} 
columnar defects formed by 5.8~GeV Pb-ions, in 
thallium-based single crystals\cite{Wahl95};  in Bi-2223 
tapes\cite{Li96} ; and in Bi-2212 single 
crystals.\cite{vanderBeek96,vanderBeek00}
    
These changes in the equilibrium 
magnetization have been attributed to magnetic 
interactions between the vortex lattice and the 
columnar defects.  By occupying a pinning site, a 
vortex gains pinning energy.  This reduction in 
system energy must exceed the energy increase 
arising from direct intervortex repulsion when a 
vortex is displaced from its natural position in 
the lattice to a particular columnar defect. For 
a defect geometry with parallel tracks that have 
a Poisson distribution of separations, Wahl et 
al.\cite{Wahl95} obtained an expression for $M_{eq}$ that 
describes reasonably well the S-shaped field 
dependence in Tl-cuprate crystals containing 
parallel CD's. 

In the Hg-cuprates investigated here, one 
might expect that the randomly oriented  columnar 
microstructure should entangle the vortices.  
Consequently it is somewhat curious that   the 
$M_{eq}$ in polycrystalline materials with random 
CD's can resemble single crystals with parallel 
defects.  We have suggested\cite{Ossandon01} that this similarity 
originates from an anisotropy-induced 
``refocussing" of the defects and field toward 
the crystalline $c$-axis.  For highly anisotropic 
single crystals, it is long recognized that only 
the normal component of field is effective.\cite{Blatter92}   
More recent studies of polycrystalline Hg-cuprates 
demonstrated a recovery of vortex 
variable range hopping (VRH), very similar to 
that observed in YBaCuO single crystals 
containing parallel CD's.\cite{ThompsonPRL97}
Among the Hg-cuprates 
with 1, 2, and 3 adjacent Cu-O layers, the 
recovery of VRH was most pronounced in the Hg-1223 
material with the largest mass anisotropy 
parameter $\gamma$. According to the theoretical 
development, the complexity of randomly oriented 
columnar defects in a polycrystalline material is 
reduced to some degree by a large superconducting 
anisotropy, restoring the simple physics of a 
crystal with parallel pins.  Thus vortex pinning 
by random CD's provides a reasonable qualitative 
explanation for the reduction in equilibrium 
magnetization, despite the complexity of the 
materials in real space.

Next we use the vortex-defect interaction 
theory of Wahl et al.\cite{Wahl95} to model our experimental 
data.  From the theory, one has 
\begin{eqnarray}
M_{eq} & = & 
       -(\varepsilon_0/2\Phi_0) \times \ln(\beta 
             H_{c2}/B) \nonumber\\
         & & -(U_0/\Phi_0)
             \left\{1 - \left[1+\frac{U_0 
B_{\Phi}}{\varepsilon_0 B}\right] \exp\left(-\frac{U_0 
B_{\Phi}}{\varepsilon_0 B}\right)\right\}
\end{eqnarray}
where $\varepsilon_0 = [\Phi_0 /4 \pi 
\lambda_{ab}]^2$ is the line energy, $U_0$ is the 
pinning energy, and $B = (H + 4 \pi M) = H$ since 
$M$ is small.  The first term is the conventional 
London expression.  The second is an added term 
to account for interactions; it is most 
significant in intermediate fields and it 
vanishes in large fields $B \gg B_{\Phi}$.  In 
modeling of data at various temperatures $T$ (77, 
90, 100, and 110~K), our objective was to 
maintain an internally consistent set of 
parameters.  The results are shown as solid lines 
in Fig. 6a for the sample with $B_{\Phi} \sim 
1$~T and in Fig. 6b for the sample with a higher 
defect density.  Given the complexity of the 
polycrystalline material and of the vortex and 
defect arrays, the description of the 
experimental data (filled symbols) is relatively 
good.  The values for the pinning energy are a 
significant fraction of the line energy and are 
quite reasonable, with $U_0 = 0.65 \varepsilon_0$ 
and $0.8 \varepsilon_0$, respectively.  To obtain 
reasonable modeling of the data for the two 
irradiated materials, we used values of 2.0 and 
2.5~T, respectively, for the effective defect 
densities $B_{\Phi}$.  The differences from the 
nominal values (calculated from the proton 
fluences) may arise from some overlap of tracks 
at the highest fluence combined with uncertainty 
in calculating the production rate for CD's.  It 
is also possible that the effective values for 
$B_{\Phi}$ compensate for other factors, such as 
residual entanglement and angular distribution in 
the CD array.  Indeed, for more highly 
anisotropic Tl-2212 materials ($\gamma > 100$, 
compared with $\gamma \sim 60$ in Hg-1223), 
similar modeling\cite{Ossandon01} was achieved with $B_{\Phi}$ 
values within 20\% of those calculated from the 
proton fluence. 

An important feature of any superconductor 
is the London penetration depth.  Figure 7 
summarizes the results for these materials in a 
plot of 1/$\lambda_{ab}^2$ versus temperature.  
Values for the virgin sample were obtained from a 
standard London analysis of the data in Fig. 5.  
An extrapolation of the Ginzburg-Landau linear 
dependence near $T_c$ (dashed line) to $T = 0$ 
yields a value $\lambda(T=0) = 174$~nm, which is 
comparable with earlier determinations.\cite{ThompsonReview98}
The values for the materials with CD's were obtained 
from the modeling procedure. All of these results 
are reasonably behaved, with 1/$\lambda_{ab}^2$ 
varying linearly with $T$ at high temperature.  
It is clear that proton irradiation increased 
$\lambda$ significantly.  This is expected, from 
two theoretical perspectives. First, conventional 
GLAG theory\cite{Orlando79} predicts that the penetration depth 
$\lambda^2$ increases as $(1+\xi_0/\ell)$ when 
the electronic mean free path $\ell$ is reduced 
by scattering, as produced by irradiation-generated 
defects from neutrons and secondary 
protons released by spallation.  Second, the 
theory of Wahl-Buzdin\cite{Wahl95} 
provides that introducing 
CD's increases the penetration depth as
\begin{equation}
  \lambda^{-2}(B_{\Phi}) = \lambda^{-2}(B_{\Phi} 
= 0) \times [1-2 \pi R^2 B_{\Phi}/\Phi_0]
\end{equation}
where $R$ is the radius of the columnar track. 
The combination of these two effects leads to the 
significant, progressive increases in $\lambda$ 
observed in Fig. 7.  To compare with Eq. 2, the 
inset of Fig. 7 shows 1/$\lambda^2$, linearly extrapolated 
to $T=0$, plotted versus defect density.  The 
open symbols denote values for $B_{\Phi}$ 
calculated from the proton fluence, while the 
closed symbols are values used for the modeling, 
as discussed above.  The dashed line in the inset 
illustrates the theoretical dependence in Eq. 2 
and shows a qualitative agreement with the data; 
its slope corresponds to a value $R \approx 
8.4$~nm, somewhat below the 11~nm value obtained 
for Tl-2212 materials.  As discussed previously,\cite{Ossandon01} 
it is likely that the CD's have a large effective 
radius due to oblique passage of ion tracks 
through CuO planes and a ``halo" of oxygen 
depletion around the central amorphized zone.\cite{Zhu93}

An interesting consequence of a larger 
London penetration depth is that the vortex line 
energy $\varepsilon_0$ decreases significantly.  
As a result, one can expect the vortex pinning 
energy of a CD and its pinning force to decrease, 
reducing its effectiveness.  This effect 
counteracts the nominally positive influence of 
increasing the defect density $B_{\Phi}$.  It is 
likely that these competing effects play a major 
role in determining the range of $B_{\Phi}$ that 
maximizes the current density $J$, as illustrated 
in the insets of Figs. 2 and 3.  This competition 
is reminiscent of the formation of (point-like) 
defects in YBa$_2$Cu$_3$O$_{7-\delta}$, where 
progressive removal of oxygen creates defects, 
but also reduces their effectiveness through 
increases in the superconductive length 
scales.\cite{Ossandon92a,Ossandon92b}

\section{Summary}
We irradiated polycrystalline 
HgBa$_2$Ca$_2$Cu$_3$O$_x$ materials with 0.8~GeV 
protons to produce randomly oriented columnar 
defects via a fission process.  Enhancements in 
vortex pinning increased the persistent current 
density, with the optimum defect density 
depending on the range of field and temperature.  
Adding random linear defects significantly 
reduced the magnitude of the equilibrium 
magnetization and changed its dependence on 
magnetic field from a simple London $\ln(H)$ form 
(as observed for the virgin material) to a more 
complex ``S-shaped" dependence.  Invoking an 
anisotropy-induced ``refocusing" of the vortex-defect 
array, we have modeled these results using 
the theory of Wahl-Buzdin that incorporates 
vortex-defect interactions.  This analysis shows 
that the addition of random columnar defects 
increases the London penetration depth markedly, 
which diminishes their effectiveness for pinning 
vortices.  Overall, however, the formation of 
randomly oriented columnar defects both enhances 
the current-carrying performance of the material 
and creates interesting superconductive systems 
for study of the interaction of vortices with 
correlated disorder.

We thank M. Paranthaman for providing the 
starting Hg-1223 materials used in this study.  
The work of JGO was supported in part by the 
Chilean FONDECYT, grant \# 1000394.  Oak Ridge 
National Laboratory is managed by UT-Battelle, 
LLC for the U.S. Department of Energy under 
contract DE-AC05-00OR22725. Los Alamos National 
Laboratory is funded by the US Department of 
Energy under contract W-7405-ENG-36. 

\section{References}

\bibliographystyle{prsty}

FIGURE CAPTIONS

Fig. 1.  The magnetization of polycrystalline Hg-1223 
materials at 60~K, versus applied magnetic 
field.  Samples were irradiated with various 
fluences $\phi_p$ of 0.8~GeV protons, as shown, 
to create randomly oriented columnar defects 
with approximate ``matching fields" 
$B_{\Phi}$.

Fig. 2.  The intra-granular persistent current density 
$J$ vs temperature for various irradiated 
materials, measured in applied magnetic field 
of 0.1~T.  The inset shows that $J$ at 100~K is 
maximized at a defect density of 0.1--0.3~T.

Fig. 3. The intra-granular persistent current density 
$J$ vs temperature for various irradiated 
materials, measured in applied magnetic field 
of 1~T.  Inset: $J$ at 60~K versus defect 
density $B_{\Phi}$ exhibits a maximum at a 
defect density of $\sim$2~T.

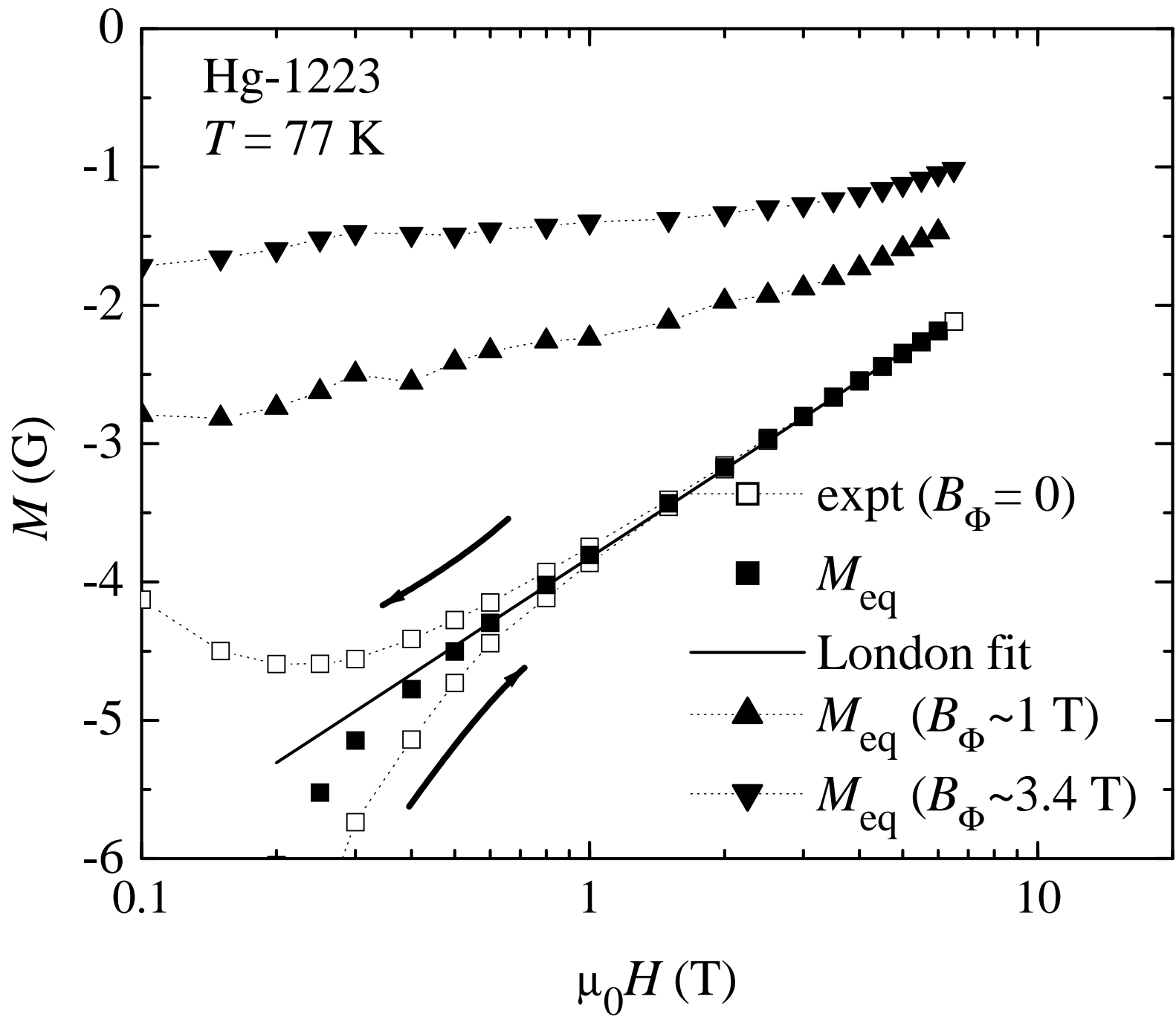
Fig. 4. The magnetization of virgin Hg-1223 at 77~K, 
versus applied magnetic field (log scale).  
Open squares show the measured $M$ in 
increasing and decreasing field history, while 
closed squares show the average $M \approx  
M_{eq}$.  The solid line is a fit to the 
conventional London $\ln(H)$ dependence.  Data 
for Hg-1223 containing randomly oriented 
columnar defects are included for comparison.

Fig. 5. The equilibrium magnetization $M_{eq}(H,T)$ 
plotted vs ln(H) for unirradiated Hg-1223 at 
the temperatures shown.  Lines are fits to 
conventional London theory. 

Fig. 6.   Plots of $M_{eq}(H,T)$ vs. $\ln(H)$ for 
irradiated Hg-1223 with (a) $B_{\Phi} \sim 1$~T 
and (b) with $B_{\Phi} \sim 3.4$~T.  Symbols 
are experimental results, while lines show 
modeling of data using Eq. 1 where effects of 
vortex-defect interactions are included.

Fig. 7.  The temperature dependence of the London 
penetration depth 1/$\lambda^2$.  Values for 
the virgin sample come from a conventional 
London analysis of the data in Fig. 5; for the 
irradiated materials, values come from modeling 
the equilibrium magnetization shown in Fig. 6.  
Inset shows values for 1/$\lambda^2$, 
extrapolated to $T=0$, plotted versus defect 
density; open symbols denote $B_{\Phi}$ values 
calculated from the proton fluence and closed 
symbols are values used in the modeling.

\end{document}